\documentclass{aa}  
\usepackage{graphicx}
\usepackage{epstopdf}
\usepackage{txfonts}
\usepackage{natbib} 
\bibpunct{(}{)}{;}{a}{}{,} 
\begin{document}

\title{Inter-network regions of the Sun at millimetre wavelengths}
% \titlerunning{}

\author{Sven Wedemeyer-B\"ohm \inst{1,2} \and Hans-G\"unter Ludwig
  \inst{3} \and Matthias Steffen \inst{4} \and Jorrit Leenaarts \inst{5}
  \and Bernd Freytag \inst{6} }
% \authorrunning{S.~Wedemeyer-B\"ohm}
   
\offprints{sven.wedemeyer@astro.uio.no}
 
\institute{Institute of Theoretical Astrophysics, University of Oslo,
  P.O. Box 1029 Blindern, N-0315 Oslo, Norway
  \and
  Kiepenheuer-Institut f\"{u}r Sonnenphysik,
  Sch\"{o}neckstra\ss e~6, 79104~Freiburg, Germany 
  \and
  Observatoire de Paris-Meudon, CIFIST/GEPI, Meudon Cedex, 92195, France
  \and 
  Astrophysikalisches Institut Potsdam, An der Sternwarte~16,
  14482~Potsdam, Germany 
  \and 
  Sterrekundig Instituut, Utrecht University, Postbus~80\, 000,
  3508~TA~Utrecht,
  The~Netherlands
  \and  
  Centre de Recherche Astronomique de Lyon - Ecole Normale SupŽrieure, 
  46, AllŽe d'Italie, F-69364 Lyon Cedex 07, France
\\
 } \date{Received date; accepted date}
% --------------------------------------------------------------------------------

\abstract{The continuum intensity at wavelengths around 1\,mm provides 
  an excellent way to probe the solar chromosphere and thus 
  valuable input for the ongoing controversy on the thermal structure and 
  the dynamics of this layer. 
} % context
{The synthetic continuum intensity maps for near-millimetre
  wavelengths presented here demonstrate the potential of 
  future observations of the small-scale structure and
  dynamics of internetwork regions on the Sun.} % aims
{The synthetic intensity / brightness temperature maps are calculated 
  on basis of three-dimensional radiation \mbox{(magneto-)}hydrodynamic 
  (MHD) simulations.  
  The assumption of local thermodynamic equilibrium (LTE) is valid for the source
  function. The electron densities are also treated in LTE for most 
  maps but also in non-LTE for a representative model snapshot.  
  Quantities like intensity contrast, intensity contribution functions, spatial
  and temporal scales are analysed in dependence on wavelength and
  heliocentric angle.} % methods
{While the millimetre continuum at 0.3\,mm originates mainly from the upper
  photosphere, the longer wavelengths considered here map the low and
  middle chromosphere. 
  The effective formation height increases generally with wavelength and 
  also from disk-centre towards the solar limb. 
  The average intensity contribution functions are usually rather 
  broad and in some cases they are even double-peaked as there are
  contributions from hot shock waves and cool post-shock regions in the 
  model chromosphere. The resulting shock-induced thermal
  structure translates to filamentary brightenings and fainter regions in
  between.  Taking into account the deviations from 
  ionisation equilibrium for hydrogen gives a less strong variation of the 
  electron density and with it of the optical depth. The result is a narrower 
  formation height range although the intensity maps still are characterised 
  by a highly complex pattern.  
  The average brightness temperature increases with wavelength and 
  towards the limb although the wavelength-dependence is reversed for 
  the MHD model and the NLTE brightness temperature maps. 
  The relative contrast depends on wavelength in the same way as the 
  average intensity but decreases towards the limb. 
  The dependence of the brightness temperature distribution on wavelength 
  and disk-position can be explained 
  with the differences in formation height and the variation of 
  temperature fluctuations with height in the model atmospheres.  
  The related spatial and temporal scales of the chromospheric pattern 
  should be accessible by future instruments. 
  } % results
{Future high-resolution millimetre arrays, such as the Atacama Large
  Millimeter Array (ALMA), will be capable of 
  directly mapping the thermal structure of the solar chromosphere. 
  Simultaneous observations at different
  wavelengths could be exploited for a tomography of the
  chromosphere, mapping its three-dimensional structure, and also for  
  tracking shock waves. 
  The new generation of millimetre arrays will be thus of great value for 
  understanding the dynamics and structure of the solar atmosphere. 
  } % conclusions

\keywords{Sun: chromosphere, radio radiation; Submillimeter;
  Hydrodynamics; Radiative transfer}

\maketitle
% 
% ================================================================================
\section{Introduction}
\label{sec:intro}

Many details concerning the small-scale structure of the solar
chromosphere remain an open issue despite the large progress during the
last decades on the observational but also on the modelling side.
Recent works suggest that the solar chromosphere within internetwork regions is a 
highly inhomogeneous and dynamic phemenon demanding for high spatial 
and temporal resolution on both sides
\citep[e.g.,][ and references therein]{2002A&A...383..283A,krijger01,ayres02,2006A&A...459L...9W,2007A&A...462..303T,2007A&A...461L...1V} 
On the numerical side, the increasing computational power but
also sophisticated new methods now allow for the necessary resolution.
On the other hand, advanced instruments enable new highly-resolved
observations of hitherto not achieved quality 
\citep[e.g., with SOT onboard the Hinode satellite, see, e.g.,][]{hinode}. 
For a detailed
comparison of observed images and three-dimensional simulations,
synthetic intensity maps need to be calculated.  Unfortunately, the
chromospheric diagnostics so far available, such as the calcium
resonance lines, require the treatment of deviations from local
thermodynamic equilibrium (LTE). A proper treatment demands for large
computational resources and sophisticated methods, rendering such
calculations hardly feasible for 3D models.  And even in case of successful
calculations, the complex translation of temperature into emergent
intensity complicates the interpretation and hampers deriving the
thermal structure from observations.

An exception, in contrast to most other diagnostics, is the radio
continuum at millimetre and sub-millimetre wavelengths. 
As its source function can be treated in LTE, it can be synthesised  
easily and hence offers a convenient way to compare observations and
numerical models.  
\citet{loukitcheva04,2006A&A...456..713L} 
took advantage of this fact
and compared a large collection of observations in the millimetre and
sub-millimetre range with synthetic brightness temperatures calculated
from models by 
\citet[][ hereafter FAL]{fal93} 
and 
\citet[][ hereafter CS]{carlsson95}. 
Recently 
\citet{2006A&A...456..713L} % --- Loukitcheva et al. (2006)
compared these models to observations done with the  
Berkeley-Illinois-Maryland Array 
\citep[BIMA,][]{2006A&A...456..697W}. % --- White et al. (2006) 
The major problem of observations at millimetre wavelengths is the 
generally poor spatial resolution which renders granular scales so far 
unaccessible -- even with the  BIMA array with its 10~antennae. 
The situation will substantially improve with the next generation of 
large interferometric arrays, e.g., the Atacama Large Millimeter 
Array (ALMA) which will be fully operational  in \mbox{$\sim 2012$} 
\citep[e.g.][]{2006SPIE.6267E...2B}. % Beasley (2006)
This instrument will provide high spatial and temporal
resolution, finally allowing to observe the small-scale structure of
the solar chromosphere in detail.  

Here we use three-dimensional radiation hydrodynamics simulations
to synthesise the continuum intensity at millimetre wavelengths 
from 0.3\,mm to 9\,mm 
\citep[see][ for a precursory study]{2005astro.ph..9747W}.
Since the simulations do only include weak magnetic fields or no field, 
the present analysis refers to internetwork regions only. 

In Sect.~\ref{sec:instruments} some instruments, that are potentially 
interesting for solar observations, are introduced. 
The numerical simulations and the method of producing synthetic 
intensity images are described in Sects.~\ref{sec:model} and 
\ref{sec:specsyn}, respectively.  
The results are presented in
Sect.~\ref{sec:results}, followed by discussion and conclusions in
Sect.~\ref{sec:discus} and Sect.~\ref{sec:conclusions}.  

% ================================================================================
\section{Instruments}
\label{sec:instruments}

The instruments addressed in this section can (optionally) be used as 
interferometers and are potentially interesting for solar observations. 
Although the angular resolution of an interferometer can be very large 
depending on the maximum baseline (connecting two individual antennae), 
there is an effective maximum resolution on which
highly reliable images of objects such as the Sun, with complicated
emission  filling  the  whole primary beam of each antenna, can be made.
Being a dynamic  object  the amount of received data constraining a snapshot image
equals twice the number of  baselines (real and imaginary parts of each 
visibility being counted separately). The number of unknowns for the 
image construction, however, is equal to the
number of  independent synthesised beam areas within the primary beam.
Reliable imaging requires that there are more data constraints than 
there are unknowns. This condition defines the effective resolution 
of reliable images for the interferometer. 
Here we estimate this resolution by assuming a homogeneous 
distribution of the synthesised beams ("image elements") 
over the primary beam area (the field of view, FOV). 
The number of those elements is equal to the number of baselines. 
An array with $N_\mathrm{a}$ antennae gives a maximum of 
\begin{equation}
\label{eq:nbaselines}
N_\mathrm{b} = \frac{N_\mathrm{a} \, (N_\mathrm{a} - 1) }{2}
\end{equation}
possible baselines. Depending on technical details of the array, only a subset 
might be realised. As such details might change, we  always refer to the 
maximum number in this work. Also note that a necessarily finite number of base 
lines limits the number of resolution elements.  
A finite number of baselines, however, can only provide an incomplete coverage 
of the \mbox{$u$-$v$}-plane (spatial Fourier space), making it difficult to determine 
the effective spatial resolution. 
For our estimate of the effective spatial resolution $\Delta \alpha$ 
we assume full \mbox{$u$-$v$} coverage in the limit of a large number 
of baselines. We finally derive the relation 
\begin{equation}
  \label{eq:instres}
  (\Delta \alpha)_\lambda  
  \approx        
  \frac{d_\lambda }{\sqrt{2\,N_\mathrm{b}}}  
  =
  \frac{d_\lambda }{\sqrt{N_\mathrm{a}\,(N_\mathrm{a}-1)}}\enspace.
\end{equation}
The primary beam size $d$ and thus the resolution, too, depends 
proportionally on wavelength $\lambda$. 
Note that our estimate is not a hard limit. 
Reliable images should be possible at significantly higher resolution, using the 
known positivity of the image and other 'a priori' constraints.   
The technique of Multi-Frequency Synthesis (MFS) uses the 
frequency-dependence of the \mbox{$u$-$v$} coverage, resulting in a larger 
number of data constraints \citep{1990MNRAS.246..490C}. 
It has thus the potential of significantly increasing the effective resolution
of reliable images. 
However, detailed imaging simulations would be needed to quantify the 
achievable resolution 
\citep{conway-privcomm07}. 

The list of interferometers in Sects.~\ref{sec:alma}-\ref{sec:rainbow}
contains characteristic properties that are relevant for the discussion in 
Sect.~\ref{sec:suggobs}. Some properties like the FOV, however, are not 
easily expressed with a single number -- in particular for the 
heterogeneous arrays CARMA, FASR, and RAINBOW since 
the different dish diameters of the individual antenna types 
result in different (wavelength-dependent) primary beam sizes. 
Please note that the description of the interferometers relies on 
information available in the literature and in the internet and is only thought to give a broad overview. 
Technical details might differ slightly in practice.

\subsection{ALMA} 
\label{sec:alma} 

Among  a large range of issues important to modern astronomy, the 
Atacama Large Millimeter Array (ALMA)  will also be used for observations of the Sun.
While its single-dish predecessor APEX (Atacama Pathfinder Experiment) 
has already been installed and delivers first scientific results, the 
construction of the array started next to the APEX site on a plateau at 
5000\,m altitude in the Chilean Andes. 
The science verification phase will most likely start in late 2009, 
followed by an early science stage from 2010, and finally full 
operation in 2012. 
All details given in this section refer to 
\citet{bastian02}, 
\citet{brown04}, 
more recently
\citet{2006SPIE.6267E...2B} % Beasley (2006)
and the ALMA web pages (e.g., \texttt{http://www.alma.info}, 
\texttt{http://www.alma.nrao.edu}, 
\texttt{http://www.eso.org/projects/alma/}). 
See also \citet{2007A&A...462..801E}. 

%-------------------------------------------------------------------------------- 
\begin{table}[b]
  \caption[h]{Frequency bands of ALMA that will be installed first 
   and corresponding  
   values for the expected primary beam size diameter $d_\lambda$ and the 
   estimated spatial resolution $(\Delta \alpha)_\lambda$ for images of 
   maximum reliability (see text for details). A total number of 50\,antennae 
   is assumed.
  }
  \label{tab:alma}
  \centering
  \begin{tabular}[b]{c|rcr|rcr|rcr|rcr}
  \hline
  \hline
  band&
  \multicolumn{3}{c}{$\nu$ [GHz]}&
  \multicolumn{3}{c}{$\lambda$ [mm]}&
  \multicolumn{3}{c}{$d_\lambda$ [\arcsec]}&
  \multicolumn{3}{c}{$(\Delta \alpha)_\lambda$  [\arcsec]} \\
  \hline
  3 & 84&\hspace*{-2mm}-\hspace*{-2mm}&116&  3.57&\hspace*{-2mm}-\hspace*{-2mm}&2.58 &  75&\hspace*{-2mm}-\hspace*{-2mm}&54& 1.51&\hspace*{-2mm}-\hspace*{-2mm}&1.10\\
  6 &211&\hspace*{-2mm}-\hspace*{-2mm}&275&  1.42&\hspace*{-2mm}-\hspace*{-2mm}&1.09 &  30&\hspace*{-2mm}-\hspace*{-2mm}&23& 0.60&\hspace*{-2mm}-\hspace*{-2mm}&0.46\\
  7 &275&\hspace*{-2mm}-\hspace*{-2mm}&373&  1.09&\hspace*{-2mm}-\hspace*{-2mm}&0.80 &  23&\hspace*{-2mm}-\hspace*{-2mm}&17& 0.46&\hspace*{-2mm}-\hspace*{-2mm}&0.34\\
  \hline
  \end{tabular}
\end{table}
%-------------------------------------------------------------------------------- 

The main array will consist of 50~antennae with a diameter of 12\,m  
in an adjustable configuration ranging from a compact size of 
150\,m to a maximum base length of $\sim 18.5$\,km. 
It is supplemented with the  (semi-independent) 
Atacama Compact Array (ACA) with 
12~antennae with 7\,m diameter and 4 12\,m-antennae.  
Each ALMA antenna will be equipped with receivers which cover 
up to ten frequency bands. 
Initially only the frequency bands in the range from 84\,GHz to 373\,GHz
will operate (see Table~\ref{tab:alma}). 
This range corresponds to wavelengths from $3.57$\,mm to $0.80$\,mm.  
The correlator can subdivide the frequency bands into a large number of spectral 
channels ($\sim 16\,000$) although lower spectral resolution can easily be 
achieved by reconfiguring the correlator or by post-correlation spectral 
averaging of the data. 
A minimum integration time of 16\,ms might be possible whereas switching between
different bands will take $\sim 1.5$\,s.

The field of view is defined by the primary beam size of an ALMA antenna, being 
\mbox{21\arcsec} at $\lambda = 1$\,mm. That is sufficient to observe the 
interior of an internetwork region. In interferometric mode ALMA will provide 
an angular resolution of \mbox{0\,\farcs015} to \mbox{1\,\farcs4}, depending on antenna 
configuration. The angular resolution corresponds to $\sim 10$\,km 
to $\sim 1000$\,km on the Sun. 
The number of baselines is 1225 for the main array with 50 large antennae 
(see Eq.~(\ref{eq:nbaselines})).
With Eq.~(\ref{eq:instres}) we estimate the effective spatial resolution 
for images of maximum reliability, $(\Delta \alpha)_\lambda$, to be of the 
order of \mbox{$\sim 0$\,\farcs42} at a wavelength of 1\,mm. 
The resolution for 0.3\,mm and 3\,mm would be \mbox{0\,\farcs13} and 
\mbox{1\,\farcs27,} resp. (see also Table~\ref{tab:alma}). 
A number of 64~antennae, as originally planned, would result in 2016 baselines. 
The total number of baselines, when including ACA, could be up to $2016 + 120$, 
reaching an effective resolution of $\sim 0$\,\farcs32 at $\lambda = 1$\,mm.

%-------------------------------------------------
\subsection{CARMA}
\label{sec:carma}

The two millimetre arrays of Owens Valley Radio Observatory (OVRO) and of 
the Berkeley-Illinois-Maryland Association (BIMA) were merged to form the 
Combined Array for Research in Millimeter-wave Astronomy (CARMA)
at Cedar Flat, California \citep{woody04, beasley03}.
This heterogeneous array consists of
six 10.4\,m antennae (OVRO) and ten 6.1\,m antennae (BIMA), resulting in 105
baselines.  There are receivers available for 115\,Ghz (2.6\,mm) and 230\,Ghz (1.3\,mm).
CARMA will be supplemented with a subarray consisting of 8 3.5\,m antennae from 
the Sunyaev-Zeldovich Array (SZA) for observations at 115\,GHz but also 35\,GHz ($
\sim 1$\,cm) with a total of 276 possible baselines. 
Different array configurations with spacing from down to 5\,m to 1.9\,km are possible. 
An angular resolution of 0\,\farcs1 can be reached (for the 230\,GHz A-array). 
See \citet{woody04}, \citet{beasley03}, and
\texttt{http://www.mmarray.org} for more information.

%-------------------------------------------------
\subsection{EVLA}
\label{sec:evla}

During the first project phase the Expanded Very Large Array 
(EVLA, see \texttt{http://www.aoc.nrao.edu/evla})
consists of the 27 VLA antennae with primary reflector diameters of 25\,m, providing 351 baselines.
In the second phase the array will be supplemented with eight new antennae, resulting in baselines 
of up to 350\,km.  The array will then have a very high angular resolution of 
0\,\farcs004 at $\lambda = 6$\,mm.  The accessible frequencies range from 1.0\,GHz to 50\,GHz, 
corresponding to 30\,cm to 6\,mm. 
The correlator will provide some thousand frequency channels.  

%-------------------------------------------------
\subsection{FASR}
\label{sec:fasr}

The Frequency-Agile Solar Radiotelescope (FASR, see 
\texttt{http://www.ovsa.njit.edu/fasr}) will combine
three types of antenna, necessary to cover the large range from
30\,MHz ($\lambda \approx 10$\,m) to 30\,GHz ($\lambda \approx 1$\,cm). 
The sub-array for the high frequencies will consist  of \mbox{$\sim $100}~antennae of 2\,m diameter 
each, setting up roughly 5000 baselines. The maximum antenna spacing will be 6\,km.  
At 30\,GHz an angular resolution of $\sim$\,0\,\farcs66 can be reached. 
FASR will produce image sequences including polarisation information with 
a time resolution of less than 0.1\,s. 
The construction is expected to be completed in 2010.   

%-------------------------------------------------
\subsection{RAINBOW}
\label{sec:rainbow}

The six transportable 10\,m-antennae of the Nobeyama Millimeter Array 
have been linked to the local 45\,m-antenna to form the RAINBOW 
interferometer (\texttt{http://www.nro.nao.ac.jp}).  
The maximum baseline length is of the order of 400\,m. 
RAINBOW can access wavelengths from 1.3\,mm to 3.5\,mm.  

% ================================================================================
\section{Numerical models} 
\label{sec:model} 

The numerical \mbox{3-D}~models used here are calculated with the radiation
hydrodynamics code \mbox{\textsf{CO$^\mathsf{5}$BOLD}} 
\citep{cobold}. 
Most of the study refers to the non-magnetic model by 
\citet[][ hereafter W04, model~A]{wedemeyer04a}, 
whereas only one snapshot is taken each from the models by 
\citet[][ hereafter S06, model~B]{schaffenberger06}  
and 
\citet[][ hereafter LW06, model~C]{leenaarts06b}.
The models are described below (see also Table~\ref{tab:models}).  

For all models a grey, i.e., frequency-independent radiative transfer 
is used. The lateral boundary conditions are periodic in all variables, 
whereas the lower boundary is ``open'' in the sense that the fluid can freely flow 
in and out of the computational domain under the condition of vanishing
total mass flux. The specific entropy of the inflowing mass is fixed to a
value previously determined so as to yield solar radiative flux at the
upper boundary. 
The upper boundary is "transmitting" in the hydrodynamics case (models A, C) 
but ``closed'' for the MHD model~B, i.e., reflecting
boundaries are applied to the vertical velocity, while stress-free conditions
are in effect for the horizontal velocities. 

Please note that we define the term {\em photosphere}  
as the layer between $0$\,km and $500$\,km in model coordinates, and the
term {\em chromosphere} as the layer above. 
The origin of the geometric height scale is chosen to
match the temporally and horizontally averaged Rosseland optical depth
unity for each model individually.   The computational time step is around 
$0.1$ to $0.2$\,s in hydrodynamics case (models A, C) and an order of magnitude 
smaller for the MHD model~B.

The utilised models still have short-comings concerning the  energy balance of 
the chromosphere but nevertheless can give a first idea of the small-scale structure and 
dynamics of this layer. Refer to Sect.~\ref{sec:disnm} 
for a discussion of the limitations of the modelling. 

%-------------------------------------------------------------------------------- 
\begin{table}[b]
  \caption[h]{Numerical models used for this study: publication
  describing the individual models in more detail, short description, 
  code used for intensity synthesis, number of time steps $n_t$, and heliocentric 
  angles ($\mu = \cos \theta$).} 
  \label{tab:models}
  \centering
  \begin{tabular}[b]{c|l|l|l|r|c}
  \hline
  \hline
model&pub.&description&synthesis&$n_t$&$\mu$\\
     &    &           & code    &      &\\
  \hline
A&W04&non-magnetic &Linfor3D&60&0.2 - 1.0\\
B&S06&magnetic,    &Linfor3D&1&1.0\\
 &   &$|B_0|=10$\,G&        & & \\
C&LW06&non-equilibrium&RH&1&1.0\\
 &   &H ionisation&        & & \\  \hline
  \end{tabular}
\end{table}
%-------------------------------------------------------------------------------- 
 
% --------------------------------------------------------------------------------
\begin{figure*}[tp]
  \centering
  \includegraphics{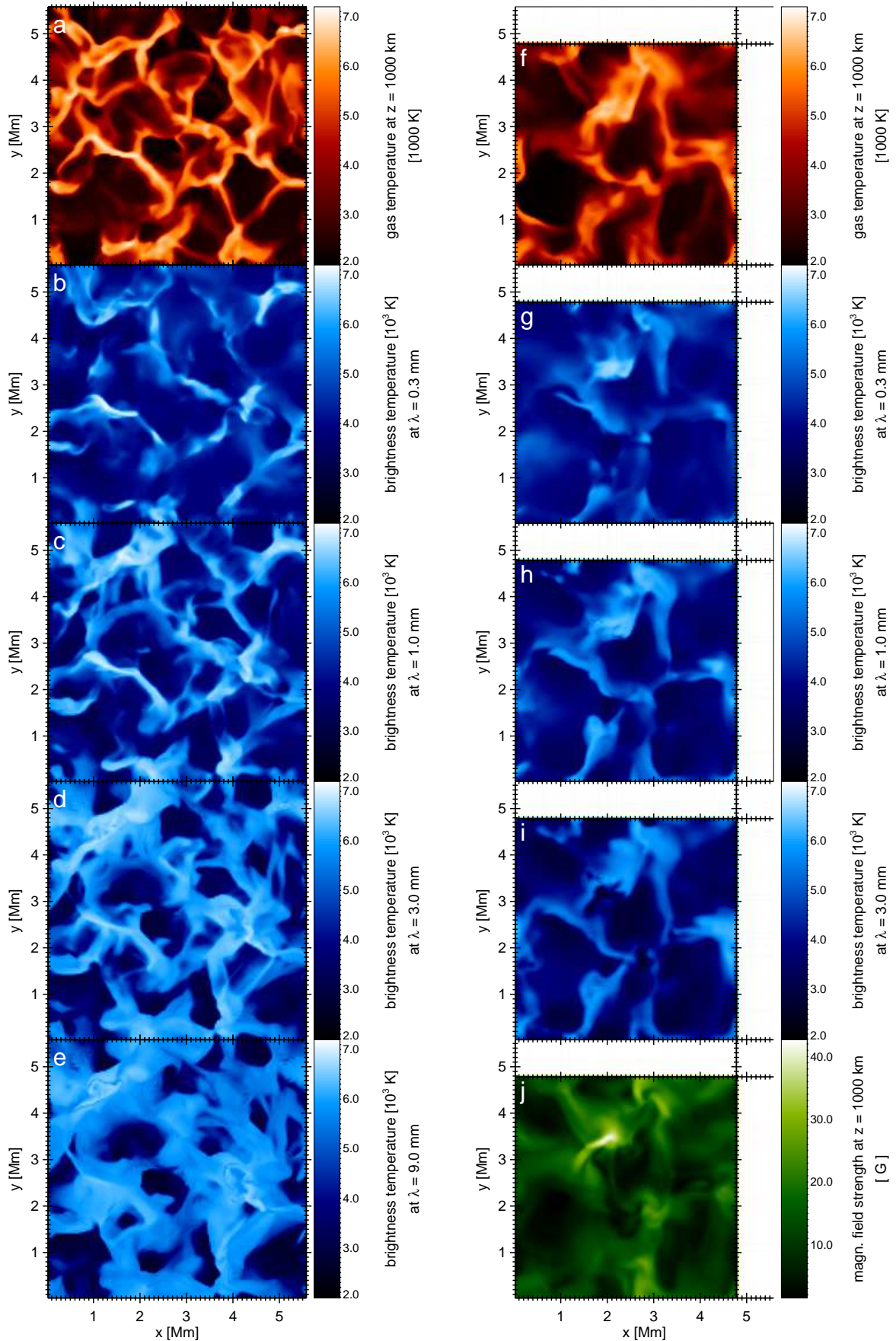}
  \caption{Horizontal maps from the non-magnetic model~A  
    (left column) and the somewhat smaller MHD model~B (right column). 
    Panels~\textbf{a)}
    and  \textbf{f)} show the gas temperature at a geometrical height of 
    $z = 1000$~km. 
    In panels \textbf{b-e)} the brightness temperature for model~A is 
    displayed for wavelengths of $0.3$~mm, $1$~mm, $3$~mm, and $9$~mm,
    respectively. 
    The brightness temperature is also shown for model~B at wavelengths 
    of \textbf{g)} $0.3$~mm, \textbf{h)} $1$~mm, and \textbf{i)} $3$~mm. 
    The magnetic field strength $|\vec{B}|$ at a height of $z = 1000$~km in model~B 
    is plotted in panel \textbf{j}.
    The quantities are colour-coded as indicated by the
    legend next to each panel. 
    Please note that the intensity contribution functions partly exceed the 
    upper boundaries of the models for the panels~\textbf{e)} and \textbf{i)}, 
    causing artefacts in the resulting brightness temperature maps. 
    For the same reason $\lambda = 9$\,mm is not shown for model~B.}
  \label{fig:intimages_a_b}
\end{figure*}
% --------------------------------------------------------------------------------
\begin{figure*}[tp]
  \centering
    \includegraphics{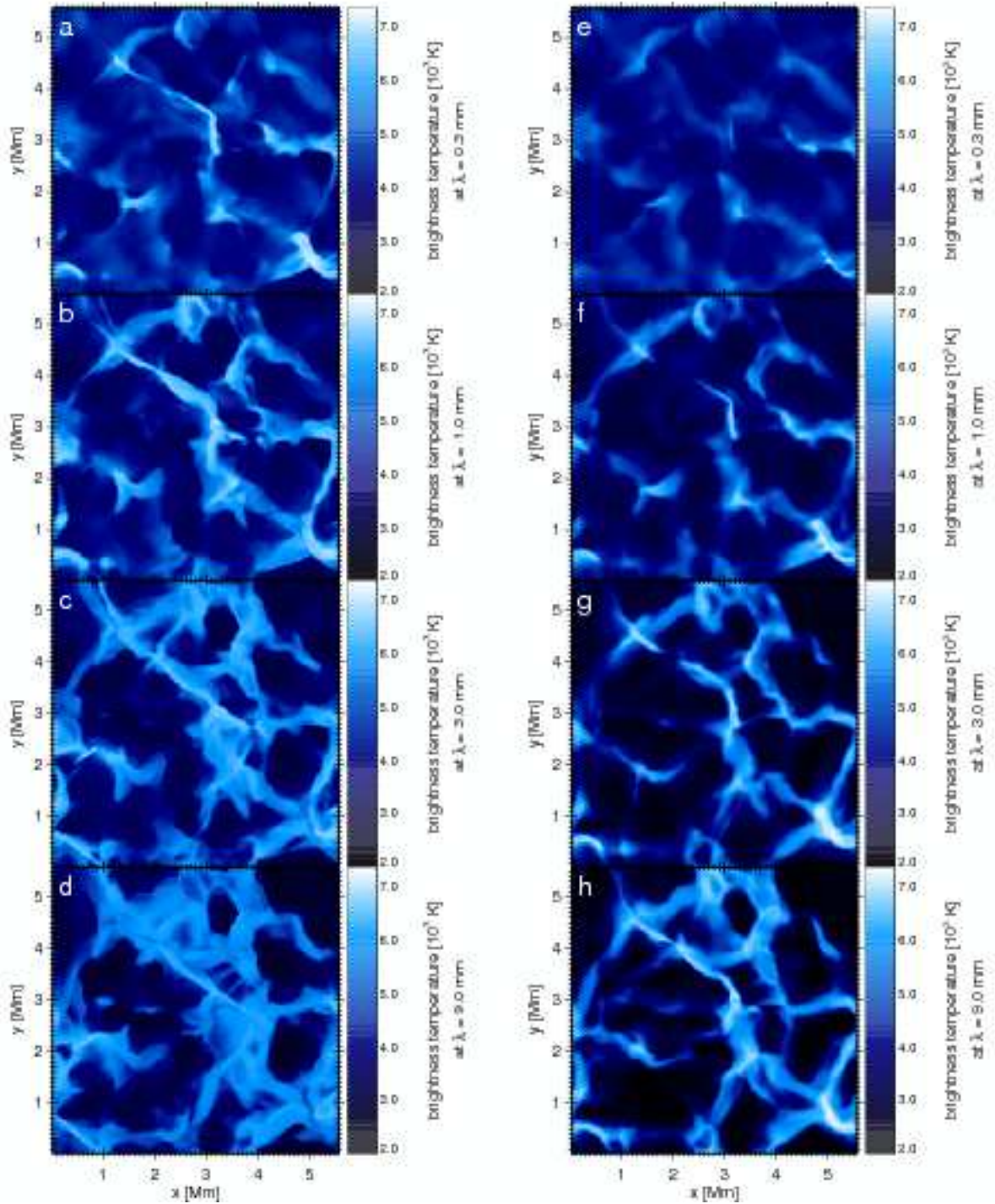}
  \caption{Horizontal maps from the non-magnetic model~C  
     with electron densities treated in LTE (left column) and 
     with the time-dependent non-equilibrium electron densities (NLTE, right column), 
     which result from the simulation. The wavelengths from top to bottom are 
     $0.3$~mm, $1$~mm, $3$~mm, and $9$~mm, respectively. 
    Please note that the average intensity contribution function for panel~d partly exceeds 
    the upper boundary of the model.
     }
  \label{fig:intimages_c}
\end{figure*}

% ================================================================================
\subsection{Field-free hydrodynamic model}
\label{sec:model_w04} 

Model A (W04) does not include magnetic fields. The computational 
domain consists of $140 \times 140$ grid cells in horizontal ($x,y$) 
and $200$ in vertical direction.
While the horizontal resolution is constantly $40$\,km, the vertical
resolution varies from $46$\,km at the bottom in the upper convection
zone at \mbox{$z = -1400$\,km} to $12$\,km for all layers above $z =
-270$\,km.  The top of the model is located at a height of $z =
1710$\,km in the middle chromosphere.  The horizontal extent is
$5600$\,km, corresponding to an angle of $7\farcs7$ in ground-based
observations.  For the analysis presented here, we use a
partial sequence with a time increment of 10\,s and a duration of
600\,s.  The model chromosphere is characterised by a mesh-like
pattern of hot shock fronts and cool post-shock regions inbetween (see
Fig.~\ref{fig:intimages_a_b}a).  

%-------------------------------------------------
\subsection{Magnetohydrodynamic model}
\label{sec:model_S06} 

Model B (S06) includes a weak magnetic field with an 
average flux density of 10\,G. The computational domain is somewhat smaller 
than for model~A. It extends over a height range of 2800\,km of which 1400\,km 
reach below the mean surface of optical depth unity (the same as for model~A) 
and 1400\,km above it. The horizontal extent is 
4800\,km~$\times$~4800\,km. With $120^3$ grid cells, the spatial resolution in 
the horizontal direction is constantly 40\,km, whereas it varies between 50 and 20\,km 
in the vertical direction, thus accounting for the varying scale heights in the 
atmosphere. 
The initial model has a homogeneous, vertical, unipolar magnetic field
with a flux density of 10\,G which is superposed on a previously computed, 
relaxed model of thermal convection very similar to model A. 
In the course of the simulation the convective flows advect the magnetic field 
towards the intergranular lanes where stronger flux concentrations ("flux 
tubes") build up. In contrast, the model chromosphere is characterised by 
a more homogeneous but more rapidly varying field distribution 
(see Fig.~\ref{fig:intimages_a_b}j). The gas temperature still is similar 
to the one in model~A (see Fig.~\ref{fig:intimages_a_b}f).

%-------------------------------------------------
\subsection{Model with non-equilibrium hydrogen ionisation}
\label{sec:model_lw06} 

Model~C 
\citep[LW06, see also][]{leenaarts06a}
is identical to model A with respect to the numerical grid, the radiative 
transfer, and the hydrodynamics solver. It also does not contain magnetic 
fields. 
The only difference compared to model~A is that the hydrogen ionisation is 
not treated in LTE but in non-equilibrium (non-LTE) by solving the 
time-dependent rate equations for a six level model atom with fixed radiative 
rates.
Hydrogen is treated as minor species so far, i.e. the non-equilibrium hydrogen 
ionisation has no back-coupling on the equation of state and on the opacities. 
The gas temperature distribution is thus (statistically) the same as for model~A.
In contrast to model~A, however, the code outputs the electron densities and 
hydrogen level populations for the LTE case and the NLTE case for each grid cell
for model~C. 
The electron density contribution of hydrogen directly results from the non-LTE 
computation, whereas the contributions of other chemical species are 
treated in LTE (see LW06 for details). 
The electron density in the model chromosphere varies much less in non-LTE compared 
to LTE. 

%================================================================================ 
\section{Intensity synthesis} 
\label{sec:specsyn} 

Continuum radiation at (sub-)mm wavelengths is mainly due to thermal 
free-free emission and originates from the chromosphere and upper 
photosphere. 
The opacity is mostly due to free-free processes (interaction of ions and 
free electrons), including H$^-$ free-free (interaction of neutral hydrogen 
atoms and free electrons). 
Owing to the large wavelength and thus small frequency the condition
\begin{equation}
  \label{eq:hnukt}
  h\nu \ll k_\mathrm{B}T_\mathrm{gas}
  \label{eq:rjcond}
\end{equation}
is fulfilled so that the Rayleigh-Jeans approximation can be used. 
According to 
\citet[][ see Eq~5.19a]{rybicki04}
and 
\citet[cf.][ p.~102]{mihalas78},  
the opacity coefficient for thermal free-free bremsstrahlung 
(ion-electron interaction) can then be written as 
\begin{equation}
\label{eq:opa}
\kappa_{\mathrm{ff},\nu} \propto
 \frac{n_\mathrm{e}\, n_\mathrm{I}}{\nu^{2}\, T_\mathrm{gas}^{3/2}}
\end{equation}
where $n_\mathrm{e}$ and $n_\mathrm{I}$ are the number densities 
of electrons and ions, resp., and $\nu$ and $T_\mathrm{gas}$ are 
frequency and gas temperature. 
In particular the dependence of $\kappa_\mathrm{ff}$ on 
$n_\mathrm{e}$ and $\nu^{-2}$ is of importance for the optical depth at wavelengths 
around 1\,mm and thus of particular interest for the study presented here. 
The free-free processes are due to collisions with electrons 
and depend on the local thermodynamic state of the electrons.  
The ratio of emission to absorption processes is thus in
local thermodynamic equilibrium (LTE), and the source function
is Planckian.
Another consequence of  Eq.~\ref{eq:hnukt} is that the contribution to 
the emergent intensity~/ brightness temperature at \mbox{(sub-)mm} 
wavelengths is linearly related to the local gas temperature in the 
contributing height range. 
The electron densities, however, can deviate from the LTE values. 
That is caused by (i)~ionisation by a non-Planckian radiation field,  in 
particular in the Balmer continuum, and (ii)~the long recombination timescales that 
hinder the hydrogen ionisation degree to follow the faster dynamic 
changes of the atmospheric conditions 
\citep[][ LW06]{carlsson02}. 
This deviation from equilibrium has an effect on the optical depth
and thus on the effective formation height of the radiation via the 
absorption coefficient (see Eq.~(\ref{eq:opa})). 
For this first qualitative study, we neglect the resulting effect on the 
opacity for the intensity synthesis from model~A and B but investigate the 
effect for model~C (see Sects.~\ref{sec:formheight} and \ref{sec:discus}).

For the radiative transfer calculations for model A and B we use 
\textsf{Linfor3D}, a 3D LTE spectrum synthesis code developed by 
M.~Steffen and H.-G.~Ludwig, which is originally based on the Kiel 
code \textsf{LINFOR}/\textsf{LINLTE}
(see \texttt{http://www.aip.de/}$\sim$\texttt{mst/linfor3D}\_\texttt{main.html}). 
Electron densities are calculated under the assumption of LTE. 
For model~C we use the RH code by 
\citet{uitenbroek00b}. 
We calculate a snapshot with LTE electron densities but also with 
the non-equilibrium electron densities, which are a direct 
result of the simulation with non-equilibrium hydrogen ionisation 
(see Sect.~\ref{sec:model_lw06}). 

Continuum intensity images are calculated at the four wavelengths 
$0.3$\,mm ($\sim 1000$\,GHz), $1$\,mm ($\sim 300$\,GHz), $3$\,mm 
($\sim 100$\,GHz), and $9$\,mm ($\sim 33$\,GHz). 
The heliocentric angle is hereafter referred to as the 
inclination angle $\theta$ and its corresponding cosine $\mu = \cos \theta$. 
For model~A we consider five different positions on the solar 
disk from $\mu = 1.0$ (disk-centre) to 
$0.2$ (near limb) with an increment of $\Delta \mu = 0.2$.
With a number of 60~snapshots this results in a total of 
$60 \times 4 \times 5$ intensity maps.  
For model~B and also the LTE and NLTE case for model~C only 
disk-centre maps for one snapshots are computed, giving 4 intensity maps 
in each case. 

For convenience, intensities $I_\lambda$ in units of 
\mbox{erg\,cm$^{-2}$\,s$^{-1}$\,\AA$^{-1}$\,sr$^{-1}$}
(as output by \textsf{LINFOR3D}) are converted to brightness 
temperature $T_\mathrm{b}$ via a relation derived from the 
Kirchhoff-Planck function in the limit of Eq.~(\ref{eq:rjcond}): 
\begin{equation}
T_\mathrm{b}= \frac{\lambda^4}{2\,k_\mathrm{B}\,c} \, I_\lambda .  
\end{equation}
Here, $\lambda$, $k_\mathrm{B}$, and $c$ are the wavelength, the
Boltzmann constant and the speed of light, respectively. 
The intensity contribution function $\mathcal{C}$ along a vertical 
view angle \mbox{($\mu = 1.0$)} is defined as 
\begin{equation}
\mathcal{C} (z) = \kappa (z) \, \rho (z) \, \mathcal{B}_\lambda (z) \, e^{-\tau_
\lambda (z)}\, .
\end{equation}
The above quantities are opacity $\kappa$ (extinction coefficient per mass
unit), density $\rho$, Planck function $\mathcal{B}_\lambda$, and optical depth $
\tau_\lambda$ which are
functions of geometrical height $z$.
Integration of the contribution functions over all heights yields the emergent 
intensity. 

Examples for the resulting intensity maps are shown in Fig.~\ref{fig:intimages_a_b} 
for models A and B at disk-centre, in Fig.~\ref{fig:intimages_c} for model~C, and 
for different inclination angles for model~A in Fig.~\ref{fig:intmuscan}. 
 
%================================================================================ 
\section{Results} 
\label{sec:results} 

%-------------------------------------------------------------------------------- 
\begin{figure}[t] 
\centering 
  \includegraphics{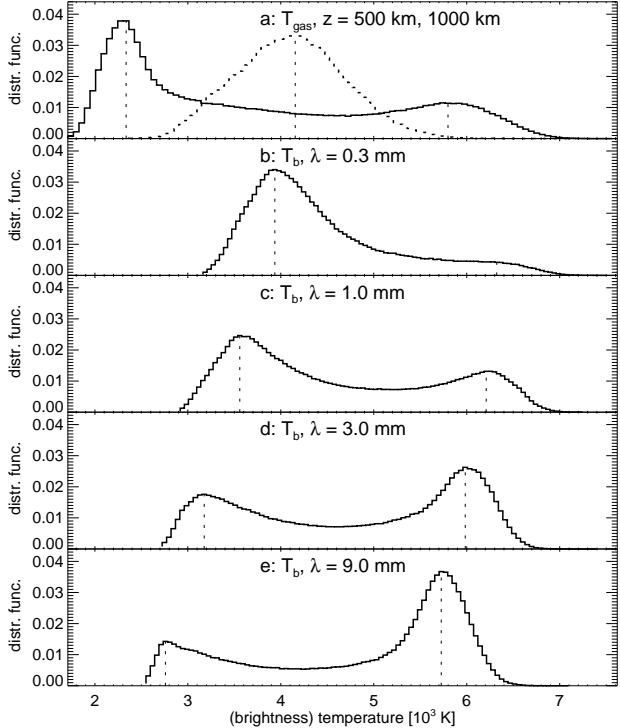}
  \caption{
    Temperature distribution in model~A: 
    Histograms for \textbf{a)} gas temperature in horizontal slices at geometrical 
    heights of $z = 500$\,km (dotted) and $z = 1000$\,km (solid), and 
    \textbf{b-e)} brightness temperature at wavelengths of $0.3$\,mm, $1$\,mm, 
    $3$\,mm, and $9$\,mm, respectively (all disk-centre). 
    All time steps of the analysed \mbox{3-D}~model sequence are taken into account. 
    The vertical dashed lines mark the positions of the individual hot and cool 
    peaks.
  } 
  \label{fig:inthist}
\end{figure} 
%-------------------------------------------------------------------------------- 

%================================================================================
\subsection{Brightness temperature distribution} 
\label{sec:distribut} 

%-------------------------------------------------------------------------------- 
\begin{table*}[t]
  \caption[h]{Average brightness temperature $\langle T_\mathrm{b} \rangle $, 
  rms variation $\delta T_\mathrm{b,rms}$, and 
  brightness temperature contrast $\delta T_\mathrm{b,rms}/\langle {T_\mathrm{b}} \rangle $ 
  for all wavelengths $\lambda$ and disk-positions $\mu$ derived from all 
  snapshots. 
  Due to the linear conversion at given $\lambda$ the values can also be interpreted 
  as intensity contrast $\delta I_\mathrm{rms}/\langle I \rangle $. 
  In some cases the corresponding average contribution function partly exceeds the upper 
  boundary of the model. The missing contribution is in the range of 0.5\,\% to 2\,\% for data 
  marked with $^*$ and $>2$\,\% for data 
  marked with $^{**}$.
  }
  \label{tab:tdist}
   \centering
  \begin{tabular}[b]{c|c|llll|llll|llll}
  \hline
  \hline
&& \multicolumn{4}{c}{$\langle T_\mathrm{b} \rangle $ [K]}&
\multicolumn{4}{c}{$\delta T_\mathrm{b,rms}$ [K]}&
\multicolumn{4}{c}{$\frac{\delta T_\mathrm{b,rms}}{\langle T_\mathrm{b} \rangle }$}\\[1mm]
  \hline
  \hline
  &&&&&&&&&&&&\\[-2mm]
model& $\mu$&
&\multicolumn{2}{c}{$\lambda$ [mm]}&&
&\multicolumn{2}{c}{$\lambda$ [mm]}&&
&\multicolumn{2}{c}{$\lambda$ [mm]}&\\
 &   & 0.3  & 1.0  & 3.0  & 9.0      & 0.3  & 1.0  & 3.0  & 9.0      & 0.3  & 1.0  & 3.0  & 9.0     \\  
 \hline
A&0.2& 5033 & 5339 & 5397    & 5332       &  820 &  726 &  642    &  587       & 0.163& 0.136& 0.119    & 0.110    \\
 &0.4& 4758 & 5145 & 5340$^*$& 5323$^*$   &  914 & 1019 &  908$^*$&  777$^*$   & 0.192& 0.198& 0.170$^*$& 0.146$^*$\\
 &0.6& 4605 & 4918 & 5149$^*$& 5178$^{**}$&  884 & 1102 & 1076$^*$&  968$^{**}$& 0.192& 0.224& 0.209$^*$& 0.187$^{**}$\\
 &0.8& 4530 & 4770 & 4983$^*$& 5031$^{**}$&  852 & 1111 & 1156$^*$& 1092$^{**}$& 0.188& 0.233& 0.232$^*$& 0.217$^{**}$\\
 &1.0& 4496 & 4684 & 4867$^*$& 4913$^{**}$&  814 & 1105 & 1188$^*$& 1155$^{**}$& 0.181& 0.236& 0.244$^*$& 0.235$^{**}$\\
\hline
  &&&&&&&&&&&&\\[-2mm]
B&1.0& 5074 & 5001 & 4860$^{**}$& 3569$^{**}$& 1380 & 1505 & 1623$^{**}$& 1213$^{**}$& 0.272& 0.301& 0.334$^{**}$& 0.340$^{**}$\\
\hline
  &&&&&&&&&&&&\\[-2mm]
C&&\multicolumn{3}{l}{LTE electron densities:}&&&&&&&&&\\
 &1.0& 4462 & 4476 & 4518$^*$& 4517$^{**}$&  718 &  998 & 1170$^*$& 1229$^{**}$& 0.161& 0.223& 0.259$^*$& 0.272$^{**}$\\
&&\multicolumn{3}{l}{NLTE electron densities:}&&&&&&&&&\\
 &1.0& 4213 & 3935 & 3768 & 3910     &  527 &  933 & 1266 & 1482     & 0.125&  0.237&  0.336& 0.379 \\
\hline
\end{tabular}
\end{table*}
%-------------------------------------------------------------------------------- 

The intensity or brightness temperature maps for the different wavelengths
all exhibit the complex pattern of hot/bright filamentary structures and 
cool/dark regions inbetween that is already seen in the gas temperature cuts
through the model chromospheres. 
See Fig.~\ref{fig:intimages_a_b} for an example of a time step from model~A
and model~B, and Fig.~\ref{fig:intimages_c}
for the LTE and NLTE cases from model~C, respectively.   
Nevertheless, there are differences in the brightness temperature distribution 
which we quantify here by means of the horizontal and, in case of model~A, 
temporal average $\langle T_\mathrm{b} \rangle $ and the relative intensity contrast 
\begin{equation}
\frac{\delta I_\mathrm{rms}}{\langle I \rangle } = 
\frac{\delta T_\mathrm{b, rms}}{\langle T_\mathrm{b} \rangle } = 
\frac{\sqrt{ \langle  T_\mathrm{b}^2 \rangle - \langle T_\mathrm{b}  \rangle ^2 }}{\langle  T_\mathrm{b}  \rangle }\enspace,
\end{equation}
where $\delta T_\mathrm{b, rms}$ is the brightness temperature fluctuation. 
The values are listed in Table~\ref{tab:tdist} for all models, wavelengths, 
and inclination angles. 
As we shall see later in more detail, the longest wavelength is only of limited 
meaning in the present study since the formation height range partly exceeds the vertical
extent of the numerical models. 

The average brightness temperature for model~A increases with  
wavelength $\lambda$ which is due to the corresponding change in effective formation height 
(see Sect.~\ref{sec:formheight}).
The same effect is found when going from disk-centre towards the limb, i.e.  
with increasing inclination angle $\theta$. 
The LTE case for model~C behaves in a similar way whereas 
the average temperatures decrease with wavelength for 
the corresponding NLTE case and also model~B. 

The absolute brightness temperature fluctuation $\delta T_\mathrm{b, rms}$ 
tends to increase with wavelength at or close to disk-centre while this trend 
reverses closer to the limb. For model~A this means an increase of the rms temperature 
amplitude from $\sim 800$\,K to $\sim 1100 - 1200$\,K. The effect is 
more pronounced for models B and C. 
The relative brightness temperature contrast  
$\delta T_\mathrm{b, rms} /\langle  T_\mathrm{b} \rangle $ depends in a similar way on 
wavelength and inclination angle. 
Despite the large temperature fluctuations in the layers apparently sampled 
by the long wavelengths (see Sect.~\ref{sec:formheight}), the contrast 
reduces to only $\sim 12$\,\% for $\lambda = 9$\,mm close to the limb. 
In contrast, the maps from model~A at disk-centre show contrasts 
of up to 24\,\% with exception 
of the shortest wavelength that provides 19\,\% at maximum. 
Model~B has much larger fluctuations with contrasts from 27\,\% to 34\,\%
implying a detectable influence of the weak magnetic field on the gas and 
brightness temperature distribution. 
The contrast in the snapshot with LTE electron densities from model~C is 
similar to model~A, although with a larger spread with wavelength.
This can be explained by the fact that the contrast for model~A relies on 
60~snapshots but only on a single one for model~C. 
The trend of the contrast to increase with wavelengt is even more pronounced 
in the NLTE case.
The more homogeneous electron densities reduce the contrast at $\lambda = 0.3$\,mm 
to only $\sim 13$\,\% but increase it for the other wavelengths. 
The different behaviour of the shortest wavelength is related to the fact that 
the bulk of emission originates predominantly from layers where the amplitudes 
of the propagating shock waves are still relatively small (upper photosphere) -- 
in contrast to the other wavelengths which sample higher layers 
(see Sect.~\ref{sec:formheight}).

The less corrugated surface of equal optical depth results in less contributions 
from layers with higher gas temperature fluctuations. 
As a consequence the NLTE case provides a cleaner measure of the layer near the 
classical temperature minimum at the top of the model photosphere where the 
temperature fluctuations are small. 
At the longer wavelengths, only the shock-patterned chromosphere is sampled in 
the NLTE case and also in LTE there are significant contributions from that layer. 
In contrast to the region of low amplitude around the classical temperature minimum
the chromosphere shows a similar pattern of hot shock 
fronts and cool post-shock regions for all heights above $z \sim 700$\,km.
In LTE, the optical depth roughly follows the temperature gradients so that the 
chromosphere is sampled on corrugated surface of equal optical depth. 
The corresponding surfaces in NLTE vary much less in height and thus cut through 
the temperature fluctuations instead of following them, explaining the higher contrast 
in NLTE compared to LTE. 
%-------------------------------------------------------------------------------- 
\begin{figure}[tp] 
\centering 
  \includegraphics{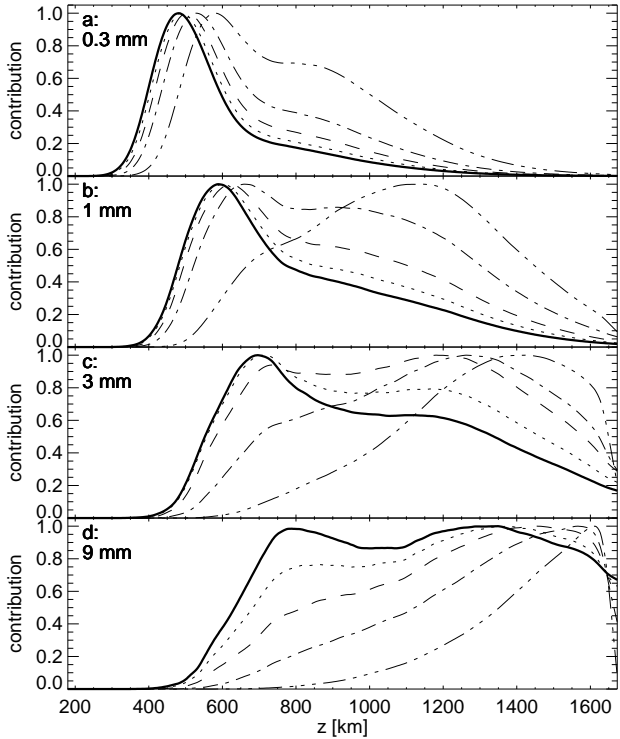}
  \caption{Normalised contribution functions for continuum intensity on the 
   geometrical height scale calculated with Linfor3D from the model by W04 
   (average over all horizontal positions and time steps in the analysed data sample)
   at wavelengths of 
   \textbf{a)} $0.3$\,mm, 
   \textbf{b)} $1$\,mm, 
   \textbf{c)} $3$\,mm, and
   \textbf{d)} $9$\,mm 
   for $\mu = 1.0$ (thick solid),    
       $\mu = 0.8$ (dotted),  
       $\mu = 0.6$ (dashed),  
       $\mu = 0.4$ (dot-dashed),  
       $\mu = 0.2$ (triple-dot-dashed).  
   The assumption of LTE was made for the calculations. 
  } 
  \label{fig:contfa}
\end{figure} 
%-------------------------------------------------------------------------------- 
\begin{figure}[tp] 
\centering 
  \includegraphics{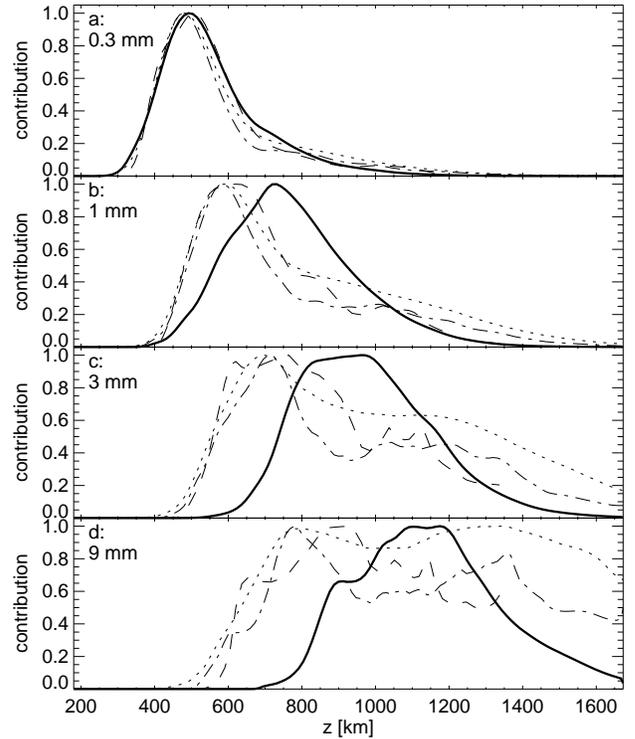}
  \caption{Normalised contribution functions for continuum intensity on the 
   geometrical height scale for average from model~A (see Fig.~\ref{fig:contfa}, here 
   dotted line), for the snapshot from MHD model~B (dashed), 
   and for the snapshot from the model with non-equilibrium hydrogen ionisation. 
   For the latter the electron densities were first calculated with RH under the 
   assumption of LTE (dot-dashed) and then with non-equilibrium electron densities 
   available from the model (thick solid). 
   The panels show different wavelengths: 
   \textbf{a)} $0.3$\,mm, 
   \textbf{b)} $1$\,mm, 
   \textbf{c)} $3$\,mm, and
   \textbf{d)} $9$\,mm. 
  } 
  \label{fig:contfbc}
\end{figure} 
%-------------------------------------------------------------------------------- 
The difference in intensity distribution is shown more detailed in form of
histograms in Fig.~\ref{fig:inthist} for the four wavelengths (all for
disk-centre only) and for gas temperature at geometric heights of $z =
500$\,km and $1000$\,km (uppermost panel) for model~A.  
As described in more detail by W04, the distribution of gas temperature 
exhibits a cool and a hot component with an intermediate range. The two 
components are due to a cool background and hot shock fronts. 
W04 also provide histograms for other heights (see Fig.~7 therein).
The histograms for continuum intensity show two components, too, although with 
varying amplitudes. At $\lambda = 0.3$\,mm a strong low-intensity component is 
visible whereas it is hard to define a high-value peak at all. 
The distribution for $\lambda = 1$\,mm is most similar to the gas temperature 
histogram at $z = 1000$\,km with peaks in almost the right proportion.
In contrast, the high-intensity peak is more pronounced than the low-value component 
in case of the longer wavelengths ($\lambda = 3$\,mm, $9$\,mm). 
The differences between the wavelengths can be understood if one considers the 
different height ranges that contribute to the continuum intensity (see 
Sect.~\ref{sec:formheight}). 
The low-value  peak  in gas temperature at $z = 1000$\,km is at lower values than 
for the brightness temperature distributions. 
A linear conversion between both temperatures due to the validity of LTE 
(see Sect.~\ref{sec:specsyn}) should produce identical distributions for gas temperature 
and corresponding brightness temperature at the same height. The differences in 
Fig.~\ref{fig:inthist}, however, illustrate the influence of the extended formation 
height ranges (see Sect.~\ref{sec:formheight}) 
which effectively mixes contributions from different layers instead of 
sampling the temperature at a fixed geometrical height, as it is the case for 
the displayed gas temperature maps.

%================================================================================
\subsection{Contribution functions and formation heights}
\label{sec:formheight}

%-------------------------------------------------------------------------------- 
\begin{table*}[t]
  \caption[h]{Heights of maximum contribution $z_\mathrm{maxc}$ (see 
  Fig.~\ref{fig:contfa}), average and variation of the heights of the 
  centroids of the spatially resolved contribution functions,  
  $\langle z_\mathrm{cc} \rangle$ and $\delta (z_\mathrm{cc})_\mathrm{rms}$, resp., 
  for all wavelengths $\lambda$ and disk-positions $\mu$ derived from all 
  snapshots. 
  The height values in square brackets only represent secondary maxima. 
  In some cases the corresponding average contribution function partly exceeds the upper 
  boundary of the model. The missing contribution is in the range of 0.5\,\% to 2\,\% for data 
  marked with $^*$ and $>2$\,\% for data 
  marked with $^{**}$.
  }
  \label{tab:zform}
   \centering
  \begin{tabular}[b]{c|c|llll|llll|llll}
  \hline
  \hline
  &&
  \multicolumn{4}{c}{$z_\mathrm{maxc}$~[km]}&
  \multicolumn{4}{c}{$\langle z_\mathrm{cc} \rangle $~[km]}&
  \multicolumn{4}{c}{$\delta (z_\mathrm{cc})_\mathrm{rms}$~[km]}\\[1mm]
  \hline
  \hline
  &&&&&&&&&&&&\\[-2mm]
model& $\mu$&
&\multicolumn{2}{c}{$\lambda$ [mm]}&&
&\multicolumn{2}{c}{$\lambda$ [mm]}&&
&\multicolumn{2}{c}{$\lambda$ [mm]}&\\
 &   & 0.3  & 1.0  & 3.0      & 9.0                 & 0.3 & 1.0 & 3.0 & 9.0 & 0.3& 1.0& 3.0& 9.0 \\  
 \hline
A&0.2& 581& 1145& 1427                & 1612                      &  818& 1076& 1137    & 1462       & 159& 231& 192& 161    \\
 &0.4& 528&  660& 1263$^*$            & 1566$^*$                  &  700&  921& 1180$^*$& 1336$^*$   & 161& 216& 254$^*$& 238$^*$\\
 &0.6& 503&  621& 1181$^*$,  [735]$^*$& 1455$^{**}$               &  645&  842& 1085$^*$& 1250$^{**}$& 156& 219& 269$^*$& 271$^{**}$\\
 &0.8& 489&  603&  713$^*$, [1173]$^*$& 1377$^{**}$, [ 867]$^{**}$&  611&  795& 1025$^*$& 1193$^{**}$& 150& 217& 274$^*$& 288$^{**}$\\
 &1.0& 481&  588&  695$^*$, [1120]$^*$& 1348$^{**}$, [ 789]$^{**}$&  588&  765&  987$^*$& 1158$^{**}$& 143& 211& 273$^*$& 297$^{**}$\\
\hline
  &&&&&&&&&&&&\\[-2mm]
B&1.0& 501&  620& 764$^{**}$, $[\sim 1040 - 1120]^{**}$&936$^{**}$       &  552&   692& 821$^{**}$&  948$^{**}$& 119& 154& 178$^{**}$& 192$^{**}$\\
\hline
  &&&&&&&&&&&&\\[-2mm]
C&&\multicolumn{3}{l}{LTE electron densities:}&&&&&&&&&\\
 &1.0& 480&  588&  720$^*$, $[\sim 1040-1200]^*$ &  780$^{**}$, [1356]$^{**}$& 659& 829&  990$^*$& 1152$^{**}$& 134& 197& 244$^*$& 271$^{**}$\\
&&\multicolumn{3}{l}{NLTE electron densities:}&&&&&&&&&\\
 &1.0& 492&  732&  960        & 1176                & 656& 868& 1074& 1255&  85& 117& 137& 164\\
\hline
\end{tabular}
\end{table*}
%-------------------------------------------------------------------------------- 
\begin{figure}[tp] 
  \centering 
  \includegraphics{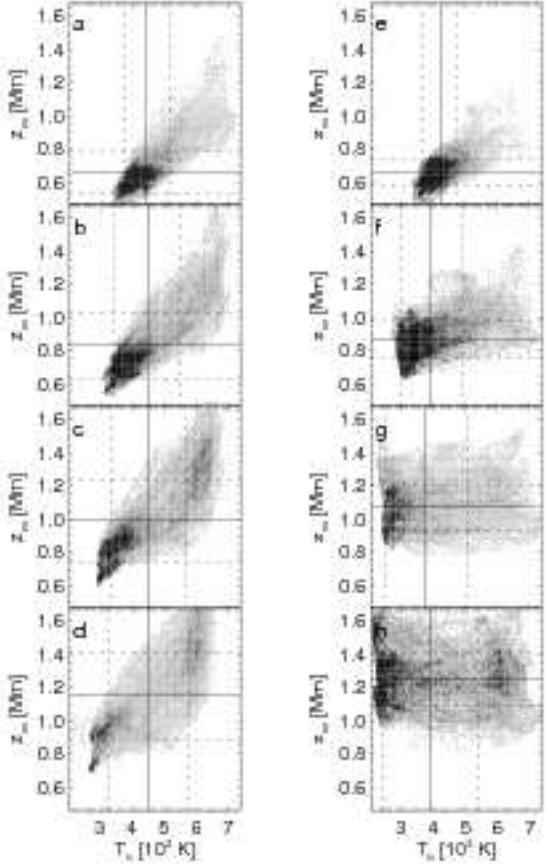}
  \caption{Formation heights $z_\mathrm{cc}$ as function of brightness temperature
  for the snapshot from model~C with LTE electron densities (left column) and 
  non-equilibrium electron densities (right column). 
  The range in brightness temperature and formation height covered by all 
  horizontal positions was divided into bins of $\Delta z = 12$\,km and 
  $\Delta T_\mathrm{b} = 50$\,K. The number of spatial positions with values 
  within the same bin (i.e., the density function) is shown as grey-scale 
  for the wavelengths of 
  $0.3$\,mm (\textbf{a},\textbf{e}), 
  $1$\,mm (\textbf{b},\textbf{f}), 
  $3$\,mm (\textbf{c},\textbf{g}), and
  $9$\,mm (\textbf{d},\textbf{h}). 
  The solid lines represent the average brightness temperatures and formation 
  heights, whereas the dotted lines mark the $1\,\sigma$ deviation.  
  }
  \label{fig:formheightmap}
\end{figure} 
%--------------------------------------------------------------------------------

In the following we describe the effective formation heights by means of 
$z_\mathrm{maxc}$, the height of the maximum of the horizontally (and 
temporally) averaged contribution functions, and by $z_\mathrm{cc}$, the 
height of the centroid of the (spatially resolved) contribution functions, 
where $z_\mathrm{cc}$ is defined as
\begin{equation}
z_\mathrm{cc} = \frac{\int z' \cdot \mathcal{C} (z')\, dz'}{\int \mathcal{C} (z')\, dz'}
\enspace.
\end{equation}
All available horizontal positions (and time steps) are used for the 
calculation of the average $\langle z_\mathrm{cc} \rangle $ and the variation 
$\delta (z_\mathrm{cc})_\mathrm{rms}$. 
The results are summarised in Table~\ref{tab:zform}.
The horizontally and temporally averaged contribution functions for 
model~A are shown in Fig.~\ref{fig:contfa} for all wavelengths and 
inclination angels. 
For disk-centre ($\mu = 1.0$), the continuum intensity at $0.3$\,mm
originates mainly from layers near the classical temperature minimum
region with a contribution peak at a height of $z_\mathrm{maxc} = 481$\,km.
Additionally, there are small contributions from the low chromosphere.
The picture is in principle the same for a wavelength of $1$\,mm,
although the peak is located somewhat higher at a height of $588$\,km
and the chromospheric contribution is larger.  Hence, these both
wavelengths map mainly the thermal structure of the top of the photosphere 
and the low chromosphere.  
Next to a maximum at $z_\mathrm{maxc} = 695$\,km the emission at 
$\lambda = 3$\,mm originates also from an extended height range in the 
chromosphere with a subtle secondary maximum at 1120\,km. 
At $\lambda = 9$\,mm all chromospheric heights contribute almost equally 
to the intensity with a maximum at 1348\,km and a less pronounced, slightly 
smaller peak (0.98) at a height of 789\,km which might be compared to the
peaks for the shorter wavelengths.  
The contribution function implies that there would still be significant 
emission from layers above the computational domain. Therefore the intensity 
images for $\lambda = 9$\,mm, as shown in Fig.~\ref{fig:intimages_a_b}, and 
also the results derived for this wavelength are incomplete to some extent
and thus of only limited significance.   
According to
\cite{loukitcheva04}, 
emission at $\lambda \ge 8$\,mm does even originate from the
transition region which is not included in the models used here.
Consequently, a more extended model would be necessary for correctly
modelling emission at long wavelengths.  

Owing to the complicated thermal structure and the corresponding 
LTE electron density, a large height range is involved in
emitting radiation at a certain wavelength, causing the contribution
function to be rather complicated.  The peaks of the individual
low-altitude components of those functions for model A (see
Fig.~\ref{fig:contfa}) move upwards by roughly 100\,km for each
increase of factor 3 in wavelength.  But additionally the
high-altitude contribution, say above $\sim 900$\,km, grows strongly
with wavelength. The mean formation height $\langle z_\mathrm{cc} \rangle$
therefore increases by $\sim 200$\,km with each factor 3 in wavelength.  

The average contribution functions for an inclined view \mbox{($\mu < 1.0$)} 
in model~A are shown in Fig.~\ref{fig:contfa}, too. At all wavelengths a
similar behaviour is obvious: The low-altitude peak (as most clearly
seen for short wavelengths at large $\mu$) moves higher up and at the
same time decreases its relative contribution until the high-altitude
range prevails. Although the latter is much broader, in contrast to
the relatively sharp low-altitude peak, a maximum can be defined in
most cases. The height of this maximum increases with decreasing
$\mu$, just as for the low-altitude contribution.
The mean formation height $\langle z_\mathrm{cc} \rangle$ 
increases by approximately 200\,km with each factor 3 in wavelength for
all inclination angles. 
 
The contribution functions for the snapshot from model~B, which are 
shown in Fig.~\ref{fig:contfbc}, are similar to those for model A, 
considering that one compares an average over 60 time steps with a 
single snapshot. Also the formation heights agree generally 
(see Table~\ref{tab:zform}), although a factor 3 in wavelength only 
increases the mean formation height by 
$\Delta \langle z_\mathrm{cc} \rangle  \sim 130$\,km
($\Delta z_\mathrm{maxc} \sim 140$\,km). 

Also the contribution functions for the snapshot from model~C 
are similar when using LTE electron densities for the intensity synthesis 
(see dot-dashed line in Fig.~\ref{fig:contfbc}). 
Using the non-equilibrium electron densities from the simulation, however, 
produces different contribution functions  for wavelengths of 1\,mm and
longer. At $\lambda = 0.3$\,mm LTE and non-LTE both are very similar 
to the other models, reflecting the fact that LTE is still a reasonable 
assumption for electron densities at that height (LW06). 
At the longer wavelengths the relevant height ranges of the contributions functions
seem to be more localized in the non-LTE case and do not exhibit such a broad range 
and double-peaked shape as for the LTE case. The difference is 
caused by the electron densities which have direct influence on the 
optical depth.
While in LTE the hydrogen ionisation and with it the electron density 
follow the strong temperature variations of the propagating shock waves, 
they do vary much less in non-LTE and tend to be at values set 
by the conditions in the shocks (LW06). The effect is that surfaces of 
constant optical depth at wavelengths around 1\,mm are much less corrugated. 
Consequently, a much smaller height range is sampled in non-LTE whereas 
in LTE the strongly varying optical depth mixes contributions from low and 
high layers, depending on the thermal structure along the line of sight 
\citep[see Fig.~3 by][]{leenaarts06b}. 

The formation heights $z_\mathrm{cc} (x,y)$ of all columns in model~C 
are shown in Fig.~\ref{fig:formheightmap} in dependence of the resulting 
brightness temperature $T_\mathrm{b} (x,y)$. 
The strong variation of the electron density and thus the optical depth in 
the LTE case makes the formation height range increase with 
brightness temperature. 
This trend is clearly visible  for all wavelengths in the left 
column of Fig.~\ref{fig:formheightmap} (LTE).
A high brightness temperature is connected to high gas temperature (in
a shock front) and thus to an increased opacity. Essentially 
the atmosphere gets optically thick already high 
up in the atmosphere if a shock wave is in the line of sight, whereas 
the cooler temperatures in the post-shock regions allow to look at much 
deeper layers. 
In NLTE (right column) this effect is less pronounced for 
the short wavelengths and essentially absent for the long wavelengths 
as the fluctuations of the non-equilibrium 
electron density are smaller. 
The spread in height is reduced in the NLTE case compared to LTE, e.g. from 
$\delta (z_\mathrm{cc})_\mathrm{rms} = 197$\,km to only 117\,km at a 
wavelength of 1\,mm. 
Still the variation prevents a strict correlation between wavelength and 
corresponding formation height that would be highly desirable for the 
derivation of the height-dependent three-dimensional thermal structure 
of the atmosphere from observations. 
Nevertheless, the (more realistic) NLTE calculations for model~C imply 
it could be possible at least in a statistical sense. 
The wavelength dependence of the average formation height remains 
very similar to the results from model~A. 
An increase by a factor of 3 in wavelength results in an increase of 
$\Delta \langle z_\mathrm{cc} \rangle \sim 190$\,km
($\Delta z_\mathrm{maxc} \sim 220$\,km). 

The clear tendency of sampling higher layers with increasing wavelength 
is expected as the opacity goes with the wavelength squared. 
In principle, the sampled height can be related to the density stratification. 
An increase of factor 3 in wavelength would then shift optical depth unity by 
that height after which the density decreased by $3^2$. In the models this height 
is mostly between 220\,km (low chromosphere) and 400\,km (towards high
chromosphere) and thus agrees well with the values directly derived from 
the average contribution functions. 

%--------------------------------------------------------------------------------
\begin{figure}[tp] 
  \centering 
  \includegraphics{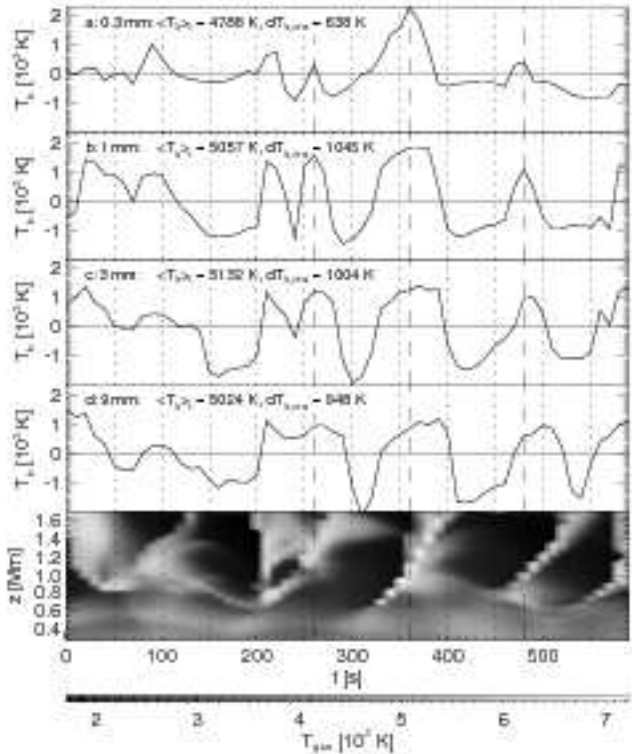}
  \caption{Temporal variation of excess brightness temperature at 
    a chosen horizontal position ($x =  4460$\,km, $y = 2020$\,km) in model~A 
    for all four wavelengths (\textbf{a}-\textbf{d}) at disk-centre. In each 
    panel the standard deviation is noted. 
    The lower panel (\textbf{e}) shows the gas temperature (colour legend below) 
    in the selected column as function of time and height. 
    The vertical long-dashed lines mark shock fronts at $\lambda = 0.3$\,mm and 
    (together with the short-dashed lines at equi-distant times) help to identify 
    the upward propagating fronts after a short delay in the brightness temperature 
    for the longer wavelengths. }
  \label{fig:brighttemp}
\end{figure} 
%--------------------------------------------------------------------------------

%================================================================================ 
\subsection{Temporal evolution} 
\label{sec:tevol}

The analysis of the temporal behaviour relies on model~A as only single 
snapshots are available for the other models. The dynamics of model~C are the 
same and also the MHD model~B is very similar in this respect. 

%--------------------------------------------------------------------------------
\begin{figure}[t] 
  \centering 
  \includegraphics{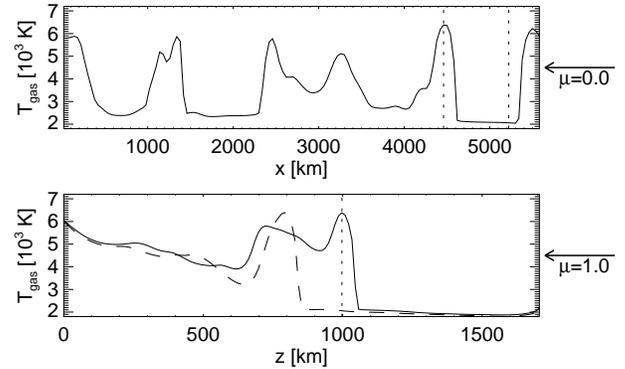}
  \caption{Gas temperature profiles along the horizontal $x$~axis (upper panel) and 
  the vertical $z$~axis for the same snapshot as in Fig.~\ref{fig:intimages_a_b} at  
  $(x,y,z) = (4460$\,km$,2020$\,km$,1000$\,km) (solid) and . 
  $(x,y,z) = (5220$\,km$,2020$\,km$,1000$\,km) (dashed). The positions
  are marked with dotted lines. The arrows at the right indicate the line of sight 
  direction at the limb ($\mu = 0.0$) and at disk-centre ($\mu = 1.0$).
  } 
  \label{fig:tempprof}
\end{figure} 
%--------------------------------------------------------------------------------
\begin{figure*}[t] 
  \sidecaption
  \includegraphics[width=12cm]{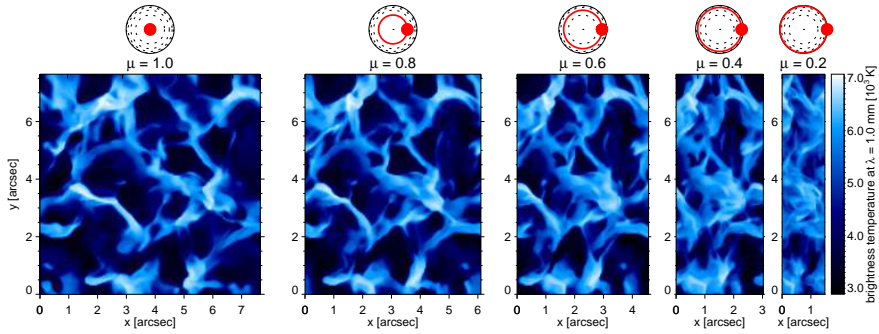}
  \caption{Variation of continuum intensity maps 
   at a wavelength of $1$\,mm with inclination angle ($\mu = \cos \theta$).  
   The thick circles with solid dot in the sketches above each panel indicate 
   the positions on the solar disk. The different panels sizes are due to
   foreshortening. 
  } 
  \label{fig:intmuscan}
\end{figure*} 
%--------------------------------------------------------------------------------

\paragraph{Pattern evolution time scale} 
The temporal evolution of the intensity pattern can be quantified in terms of 
a time in which the autocorrelation decays to the fraction $1/e$ as it has been 
done in W04 for gas temperature. Here, we apply the same procedure to time sequences 
of brightness temperature for all wavelengths and all inclination angles 
from model~A. 
The time scales are mostly around 23\,s to 24\,s for $\mu > 0.2$ and go down 
to 19\,s for $\mu = 0.2$ and $\lambda = 3$\,mm. 
The extreme case of 17.4\,s for $\mu = 0.2$ and $\lambda = 9$\,mm must regarded 
as uncertain as it is influenced by possible artefacts due the only partially 
included formation height range.   
The gas temperature in the model chromosphere itself exhibits time scales of the 
same order ($20$\,s to $25$\,s)
and stays almost constant throughout the chromosphere with a tendency towards 
higher values for the lower layers (see W04). 
Hence, it is not surprising that the intensity image sequences reproduce 
roughly the same time scales for the different wavelengths despite the different 
contribution functions.

\paragraph{Brightness temperature variations} 
The dynamic nature of the model chromosphere is very obvious if one
looks at the temporal variation of the brightness temperature in
Fig.~\ref{fig:brighttemp} (see also Fig.~10 of W04 for the chromospheric 
gas temperature variation). The amplitude of the fluctuations,	
expressed quantitatively by means of $T_\mathrm{b,rms}$ (see Table~\ref{tab:tdist},
 compare with Fig.~9 in W04), is generally smaller in the lower layers than
further up, since the shock waves steepen on their way upwards into the 
thinner atmospheric layers. 
The propagating wave fronts are clearly visible as bright streaks in the lower 
panel which shows the gas temperature in the selected column as a function of 
height and time. 
The other temperature enhancements are caused by interaction with neighbouring 
wave fronts and the fact that the waves do not only move vertically (inside the 
column) but also laterally and thus can leave and enter the selected column sideways.

Since the continuum intensity at $\lambda = 0.3$\,mm emerges from the high
photosphere and low chromosphere, it exhibits smaller variations of the
related brightness temperature compared to the remaining wavelengths
which sample higher layers.  For the particular example in
Fig.~\ref{fig:brighttemp} we find a standard deviation of 
$\sim 640$\,K at $\lambda = 0.3$\,mm and around 1000\,K for the longer
wavelengths.  
The vertical dashed lines in the figure mark shock fronts at \mbox{$\lambda = 0.3$\,mm,} 
which are formed low in the atmosphere (see lower panel). 
The longer wavelengths show an excess in brightness temperature shortly 
afterwards. 
The time difference with respect to 0.3\,mm increases with wavelength and thus 
with formation height, clearly indicating upward propagating shock waves.

The vertical gas temperature stratification for the selected column is shown 
in Fig.~\ref{fig:tempprof} at a time of 480\,s. 
The shock wave at a height of $z = 1000$\,km appears as brightening 
at a wavelength of 1\,mm at $t = 480$\,s (Fig.~\ref{fig:brighttemp}) and 
shortly afterwards at the longer wavelengths.
The absolute temperature amplitudes of the shock waves, however,  depend on 
details of the numerical modelling of the chromosphere (see discussion in 
Sect.~\ref{sec:disnm}). The determination of brightness temperatures from 
observations will thus provide an important test for the numerical models. 

%================================================================================ 
\subsection{Centre-to-limb variation} 
\label{sec:varmu} 

In Fig.~\ref{fig:intmuscan} a sequence of intensity imagesfor $\lambda = 1$\,mm with 
increasing inclination angle $\theta$ is
shown, illustrating the changes as one goes from disk-centre towards
close to the limb.  At disk-centre ($\mu = 1.0$) the chromospheric
small-scale structure with its bright filaments and dark intermediate
regions is best visible, resulting in the highest intensity contrast
(see Table~\ref{tab:tdist}). But towards the limb the contrast
decreases since the dark regions become less dominant.  This can be
explained if one considers that at disk-centre the chromosphere is
seen from above and at (or close to) the limb from the side, making
visible different geometrical aspects of the thermal structure of the
model chromosphere.  From above the model chromosphere appears as
alternating pattern of hot and cool gas (see 
Fig.~\ref{fig:tempprof}) and thus bright and dark in continuum intensity.
There are usually only one or two (or none) shock fronts in the
line-of-sight, allowing to look deep into the intermediate dark
regions.  In contrast from the side (small $\mu$) many ``propagation channels'' of
shock waves overlap along the line-of-sight, resulting in much lower
intensity contrast (see Fig.~\ref{fig:tempprof}).  Observations of the centre-to-limb 
variation of the intensity pattern may thus give valuable information on the
three-dimensional structure of the solar chromosphere.

%-------------------------------------------------------------------------------- 
\begin{figure}[h] 
  \centering 
  \includegraphics{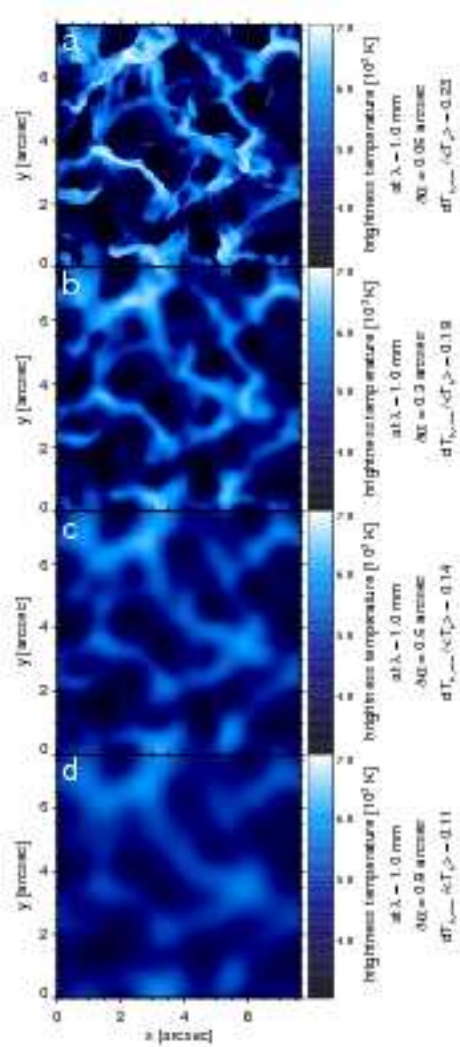}
  \caption{Intensity image at $\lambda = 1$\,mm from a single snapshot of model~A 
   at different spatial resolutions: 
   \textbf{a:} $0\farcs06$ (size of computational grid cells in the original model), 
   \textbf{b:} $0\farcs3$, 
   \textbf{c:} $0\farcs6$, 
   \textbf{d:} $0\farcs9$. 
   The value range is the same for all panels. The brightness temperature 
   contrast is noted right next to each panel.
  } 
  \label{fig:resimg}
\end{figure} 
%-------------------------------------------------------------------------------- 

%================================================================================ 
\subsection{Spatial resolution} 
\label{sec:resolution}

A major disadvantage of past and present instruments for the (sub-) millimetre 
range is an only poor spatial resolution which hindered a detailed study 
of the small-scale structure of the solar atmosphere. 
In order to estimate the spatial resolution which is needed to resolve 
the pattern clearly present in the model chromospheres used here,  
we choose a representative intensity image from model~A for
$\lambda = 1$\,mm and degrade  it  artificially by convolution with a Gaussian kernel,
mimicking instrumental properties.  The result is shown in
Fig.~\ref{fig:resimg} for different widths (FWHM) of the Gaussian.
The finest structures are already hardly visible at a resolution of
\mbox{0\,\farcs3} whereas the larger mesh-like pattern is still discernible at
\mbox{0\,\farcs9}. 
The brightness temperature contrast, which is noted next to the 
panels in Fig.~\ref{fig:resimg}, reduces with spatial resolution 
$\Delta \alpha$. 
This trend can be fitted with
\begin{equation} 
\frac{\delta T_\mathrm{b, rms}}{\langle T_\mathrm{b} \rangle } \propto 
e^{-\Delta \alpha / D}
\enspace, 
\end{equation}
where $D$ is the characteristic length scale of the chromospheric pattern
(derived from the fit).
This dependence is similar for all wavelengths and inclination angles 
although the length scale for the best fit is different. 
Except for $\mu = 0.2$, where $D \approx 290$\,km for all wavelengths, 
the length scale increases with wavelength. 
At disk-centre it increases by $\sim 100$\,km 
for a factor three in wavelength from  
730\,km at $\lambda = 0.3$\,mm to 
1030\,km at $\lambda = 9$\,mm. 
This can be explained by the fact that the propagating shock waves, which 
produce the pattern, only become visible at the bottom of the chromosphere, 
whereas in the higher layers the interaction of neighbouring wave fronts 
developed into a more distinctive pattern of larger mesh-size (see Fig.~2 
by W04). 
On the other hand, $D$ decreases with increasing inclination angle as 
one looks at the mesh-like pattern from above at disk-centre but 
sideways on the shock fronts close to the limb 
(see Sect.~\ref{sec:varmu}   and Fig.~\ref{fig:intmuscan})

%================================================================================ 
\section{Discussion} 
\label{sec:discus} 

\subsection{Numerical modelling}
\label{sec:disnm}
 
The results of this study demonstrate the constraints on spatial and
temporal resolution for observations of the solar inter-network 
chromosphere with future \mbox{(sub-)}millimetre instruments but  
should be regarded qualitatively only since some important ingredients
are still missing in the models used here. 
The example of model~B implies that even weak magnetic fields, 
as they are expected for solar internetwork regions,  
will have most likely important (indirect) influence on structure and dynamics 
of the chromosphere. Nevertheless, the qualitative picture 
of pronounced inhomogeneities clearly present in the non-magnetic model~A 
are also found for the weakly magnetic model~B (see Sect.~\ref{sec:discmhd}). 

A more severe limitation concerns the assumption of LTE for the radiative 
transfer and the equation of state, which is made here 
in order to keep the simulations feasible. At chromospheric heights the 
assumption is no longer valid. Instead a by far more involved 
treatment of non-equilibrium and radiative non-LTE effects  is required
(e.g. hydrogen ionisation). 
This will certainly influence the energy balance and with it the
thermal structure of the chromosphere. 
Taking into account the time-dependence of hydrogen ionisation, which is 
not instantaneous at chromospheric heights but rather is governed by long 
recombination time scales, leads to deviations from ionisation equilibrium 
and thus electron densities that can differ significantly from the
corresponding equilibrium values
\citep[][ LW06]{carlsson02}. 
The effect on the optical depth at millimetre wavelengths is significant, 
The electron density affects the optical depth and thus determines which
layer is effectively mapped. 
From the analysis of a recent simulations that accounts for non-equilibrium 
hydrogen ionisation (model~C, see LW06) we
learn that (i) the height of optical depth unity at a wavelength of
1\,mm is on average close to the LTE case and does not vary that
strongly, (ii) the intensity fluctuations and the average brightness
temperature are reduced, (iii) the resulting intensity maps are still
qualitatively similar (see Sect.~\ref{sec:distribut}). 
However, the back-coupling of the non-equilibrium on the equation of state 
and on the opacities -- a final step that has to be done -- will most likely 
increase the peak temperatures of the shock waves.

\subsection{Magnetic field}
\label{sec:discmhd}

The snapshot from magnetohydrodynamic model~B  gives a first idea about the 
influence of the weak chromospheric magnetic field when compared to the 
non-magnetic model~A (for $\lambda < 9$\,mm). 
Already the brightness temperature maps in Fig.~\ref{fig:intimages_a_b} 
show a similar pattern for both models. The characteristic length scale 
of the mesh-like brightenings, which are due to hot shock fronts, seems to 
be larger in the MHD case. 
The formation heights are very similar and are only slightly higher in 
model~B although this might be due to the particular snapshot chosen. 
The average brightness temperature is higher in model~B for $\lambda = 0.3$\,mm 
and 1\,mm but is the same than for model~A at 3\,mm. The average actually 
decreases with wavelength for model B whereas it increases for model~A. 
The rms brightness temperature variations increase with $\lambda$ in both cases
but are 400\,-\,500\,K higher in the MHD model. 
The different behaviour of the two models might be explained with the finding 
that the magnetic field -- although still quite weak -- becomes more important 
with increasing height (S06) and might influence the temperature distribution 
by suppressing power of the propagating waves. Also the field concentrations 
in the photosphere might change the wave excitation which should also 
become visible in the higher layers. 
The comparison of magnetic and non-magnetic models should be repeated in more 
detail and with a larger number of snapshots in the future, but alreadya
this first qualitative study suggests that high-resolution 
\mbox{(sub-)}millimeter observations can give (indirect) information on 
the magnetic field topology of the solar chromosphere. 

\subsection{Synthetic observations}

The computation of synthetic images that can be directly compared 
to observations requires more than a simple convolution with a Gaussian 
(Sect.~\ref{sec:resolution}), which is strictly speaking only valid in 
the limit of full \mbox{$u$-$v$}-coverage (see Sect.~\ref{sec:instruments}).
For more realistic maps one needs to convolve a
synthetic image with the primary beam response of the antenna used, 
Fourier-transform the result, and multiply with the \mbox{$u$-$v$}-coverage of the 
array configuration. The Fourier back-transform yields then the 
"dirty map". Such a detailed image construction, of course, depends on 
specific instrumental properties and array configuration at the time 
of the observation. Nevertheless, synthetic images can be easily generated 
to match the properties of a particular observation. 

\subsection{Comparison to observations and other models.} 

\citet{loukitcheva04} compare the dynamic 1D simulation by CS and the 
semi-empirical FAL model~A with a large number of observations (see references 
therein). 
They conclude that the FAL model A could also account for the measured brightness
temperatures, at least when combined with the CS model. 
Although the brightness temperatures calculated from the FAL~A stratification 
are still within the error bars of the observations, the values tend to be at the 
upper limit if not even exceeding it. 
The CS model matches the observations much better as it predicts much lower 
temperatures than FAL~A. The discrepancy is still small for short wavelengths 
($\lambda \sim 0.1$\,mm), 
where the intensity originates from deep in the atmosphere, but increase to the order 
of 3000\,K for the longer wavelengths. 

The average brightness temperatures in our models is similar to the values for the 
CS simulation and the observations for the shorter wavelengths 
($\lambda \leq 1$\,mm). 
For the longer wavelengths, however, CS, FAL~A, and also the observed 
temperatures increase strongly with wavelength whereas our models yield 
considerably smaller average values of the order 4000\,K~-~5000\,K. 
In case of $\lambda = 9$\,mm the discrepancy can partly be attributed to 
the fact that the continuum forming layers exceed the vertical extent 
of model A and B and also C in the LTE case although this argument does not hold 
for the NLTE case for model C. The transition region and its steep temperature 
gradient, which is included in the CS and FAL~A models but not in ours, is a more 
likely source of discrepancy. In the CS simulation the transition region 
starts around a height of 1500\,km from where in our models a significant 
part of the emergent continuum intensity at long wavelengths originates. 
Consequently the temperatures at long wavelengths would be too low and increase 
of the average with wavelength not as steep as found by CS and FAL. 
Last but not least the brightness temperatures distribution is affected by 
uncertainties in the chromospheric energy balance due to a too simple radiative 
transfer and the still missing back-coupling of the non-equilibrium hydrogen 
ionisation to the equation of state. The latter would lead to higher shock peak 
temperatures. 

The brightness temperature variations derived in this work 
(see Table~\ref{tab:tdist}) tend to increase with wavelength (at disk-centre) 
and thus show the same tendency as found by  
\citet{loukitcheva04}.  
They also explain this behaviour with different formation height ranges 
increasing with wavelength.
For the very long wavelengths ($\lambda > 8$\,mm), however, the temperature 
variations in the CS model decrease again. The rms variation,  shown in Fig.~4 of 
\citet{2006A&A...456..713L} as function of wavelength, actually has a 
maximum at around 2\,mm with $\delta T_\mathrm{b, rms} = 930$\,K. 
The corresponding variations in the models presented here are significantly 
larger for all considered wavelengths. 
This is related to the differences in the profile and propagation of the
shock waves in the 1D model CS and our 3D models. 
While sawtooth-like shock profiles are clearly present in the CS model
(see Fig.~3 by 
\citeauthor{loukitcheva04}, \citeyear{loukitcheva04}, and 
Fig.~1 by
\citeauthor{2006A&A...456..713L}, \citeyear{2006A&A...456..713L}), they 
are not so easily visible in the models used here
(see Fig.~\ref{fig:brighttemp}). In 3D -- in contrast to a 
1D model -- shock waves propagate in all spatial directions and interact 
with neighbouring wave fronts and this way obscure the shock signatures in 
vertical columns. 
Nevertheless, there are excess temperatures of more than 1000\,K in both 
kinds of model.   
The case of 0.05\,mm for the CS simulation varies less and resembles more 
a regular oscillatory behavior as it originates from further down in the 
photosphere. 

\subsection{Formation heights}

Double-peaked contribution functions were found by 
\cite{loukitcheva04} 
for simulated brightness temperature at millimetre wavelengths for
single snapshots from the 1D simulation by CS 
(see Fig. 5 therein). A secondary maximum is 
also visible in the average contribution for $\lambda = 1$\,mm.  
This agrees well with the contribution functions for the models 
used here. Secondary maxima are more obvious in the contribution 
functions for individual columns, whereas the functions shown 
in Figs.~\ref{fig:contfa} and \ref{fig:contfbc} are smoother as 
they represent averages from 3D calculations with a large number 
of columns. 
More than one peak appears if there is more than one shock wave 
in the line of sight or at least a shock wave above, i.e. at smaller 
optical depth,  the average formation height range set by the 
background stratification. Multiple contribution peaks are thus 
a direct imprint of the highly dynamic simulations and 
details certainly depend on the properties of the shock waves. 

A comparison of the average contribution functions for the 
models by CS and FAL~A 
\citep[Figs.~6 and 7 in][ resp.]{loukitcheva04} 
shows that the formation heights are in line with our results 
with exception of $\lambda = 10$\,mm in FAL~A which is formed 
at a height of $\sim 2250$\,km in the transition region. 
In contrast we find good agreement for the longest wavelength 
in our models with the CS model. For the other wavelengths (0.3\,mm - 3\,mm), 
however, the average formation height tends to be slightly lower 
in our models compared to CS and 
FAL~A. 
According to \citet{loukitcheva04}, the emission at $\lambda \ge 8$\,mm does 
originate from the transition region which is not included in the 
models used here. Consequently, a more extended model would be necessary 
for correctly modelling emission at long wavelengths.  

The formation height range for model~A seems to be too broad to
derive information about the thermal structure from observations. 
The situation improves when taking into account the non-equilibrium 
electron density contribution due to hydrogen as this flattens the 
surface of equal optical depth. 
As the electrons in chromosphere are predominantly due to hydrogen,
non-equilibrium modelling of other electron donators (which are currently 
treated in LTE) are thus expected to have only a minor influence on the 
electron number density. 
But even with an advanced non-equilibrium modelling the contribution 
functions extend over a significant height range, causing the effective 
formation heights to vary over $\sim 100$\,km (model~C, see Table~\ref{tab:zform}). 
This is illustrated very clearly in Fig.~3 by 
\citet{leenaarts06b}. 

The resulting brightness temperatures should thus be interpreted very carefully
as they represent the integrated physical state of an
extended height range. Hence, a brightness temperature can deviate
significantly from gas temperature at a geometric height in the case
of an inhomogeneous and rapidly evolving chromosphere.

Nevertheless, our study is promising in the sense that simultaneous 
observations at many millimeter wavelengths can still serve as 
a statistical tomography of the chromospheric layers. Not only the 
wavelength dependence but also the centre-to-limb variation of the 
formation height might prove useful in this respect.

\subsection{Suggested observations}
\label{sec:suggobs}

Based on the results concerning the spatial resolution and effective 
formation heights discussed above, we can suggest which will be the 
most promising instruments (see Sect.~\ref{sec:instruments}) for 
observing the small-scale structure of the solar chromosphere. 

Although RAINBOW does access the wavelength range of interest here, 
the small number of antennae will not allow for the high spatial 
resolution required for imaging the chromospheric fine-structure. 
For the same reasons CARMA cannot provide brightness temperature maps that 
have high spatial {\em and} temporal resolution at the same time. 
EVLA does consist of more antennae, resulting in a much larger number of 
baselines and thus a much higher spatial resolution. 
The minimim wavelength of 7\,mm, however, is at the limit 
for our application as a significant part of the continuum radiation 
at this wavelength and beyond originates from layers that are not or 
only partly within the computational domain of our models. 
Nevertheless, EVLA is certainly of interest for observing the 
upper chromosphere / transition region. In this respect it connects to 
the wavelength range covered by ALMA. 
The shortest accessible wavelength of FASR will be even 1\,cm. 
Just as EVLA, it won't be able to map the low and middle chromosphere but
will be of great value for investigating the solar corona at even higher 
spatial resolution than EVLA.  

The by far most promising tool for imaging the chromospheric 
fine-structure at high cadence is ALMA as it can access the spatial 
and temporal scales implied by our study. 
The effective spatial resolution, however, might be critical for the longer 
wavelengths.
A possible solution could be the application of Multi-Frequency Synthesis 
(MFS, see Sect.~\ref{sec:instruments}). 

The wavelength range of the individual bands of ALMA 
(see Sec.~\ref{sec:alma}) corresponds to an expected difference 
in formation height range of $\sim 100$\,km between lower 
and upper limit of the band. 
The thousands of spectral channels by ALMA, which are 
output simultaneously for each wavelength / frequency band, thus 
correspond to different layers, stacked with slightly differing formation  
heights. Tomographic reconstruction methods might then be applied to 
derive the three-dimensional structure of the layer sampled by the band. 

Simultaneous observations in different frequency bands will not be possible
but switching between bands can be as fast as 1.5\,s which is fast 
enough in case of the solar chromosphere. 
As upwards propagating waves show up in the different bands subsequently 
with increasing wavelength (see Sect.~\ref{sec:tevol}), it would be best to 
cycle through the different bands from high to low frequencies. 
Tracking waves through the atmosphere should be possible with ALMA. 
The subsequent appearance of particular features at the different 
wavelengths will enable the determination of the propagation speed and 
might -- in combination with measured amplitudes -- give further insight 
into wave dissipation and with that into the acoustic heating contribution. 
Combined simultaneous observation runs with optical diagnostics for the low 
and middle photosphere have the potential to reveal the excitation sources 
of waves. 

%================================================================================ 
\section{Conclusions} 
\label{sec:conclusions} 

Based on recent three-dimensional (magneto-)hydrodynamic simulations we conclude that 
future (sub-)millimetre instruments like ALMA will be capable of linearly
mapping the gas temperature distribution in the chromosphere. 
The temporal and spatial resolution of ALMA will be sufficient to resolve structure 
and dynamics on scales inherent in the models presented here.

The inhomogeneous and dynamic nature of the (model) chromosphere causes   
the intensity contribution functions and with that the formation height ranges 
to be extended and to vary in time and space. 
The variations, however, stay moderate as the electron density, which mainly defines 
the optical depth at wavelengths around 1\,mm, is connected to the rather 
well-stratified ionisation degree of hydrogen.
The formation height range increases with increasing wavelength. 
The same effect is found for decreasing inclination angle, i.e. higher layers are 
sampled close to the solar limb than at disk-centre. 
This behaviour could be exploited for a (statistical) tomography of the thermal 
structure of the solar atmosphere. 
In the case of ALMA the upper photosphere to middle chromosphere will be mapped 
with the different wavelength bands with high spatial and temporal resolution.  
The large number of spectral channels within each wavelength band will result in 
many simultaneous maps which correspond to continuously stacked formation heights. 
Tomographic reconstruction techniques might then allow {\em volume imaging} of the solar 
atmosphere. 

Hence, it should even be possible to track propagating hydrodynamical waves in the 
solar chromosphere and to study chromospheric oscillations in more detail. 
This will have import implications for the discussion of the heating mechanism not only for the 
Sun but for stellar chromospheres in general. 
The range of application is broad and does include the study of the fine-structure 
of the magnetic network, too, direct measurement of the magnetic field topology, 
and even full-disk mosaics. 
Thus, ALMA will be of great value for finally resolving the discussion about the 
dynamics and the true structure of the solar chromosphere.

%================================================================================ 
\begin{acknowledgements} 
  In memory of Prof.~Dr.~H.~Holweger. 
  The authors thank (in alphabetical order) 
  T.~Ayres, T.~Bastian, J.~Bruls, M.~Carlsson, J.~Conway, M.~Loukitcheva, 
  R.~Osten,  O.~Steiner, and S.~White for helpful
  comments and discussion.  
  SWB also likes to thank the organisers and participants 
  of the APEX/ALMA meeting in Lund, 2007.
  SWB was supported by the Research Council
  of Norway, grant 146467/420, and the {\em Deutsche
  Forschungs\-gemein\-schaft (DFG)}, project Ste~615/5.
\end{acknowledgements} 
%================================================================================ 
\bibliographystyle{aa}

%================================================================================ 
\end{document}